\documentclass[3p,times]{elsarticle}

\usepackage{ecrc}

\usepackage{epstopdf}

\volume{00}

\firstpage{1}

\journalname{Nuclear Physics A}

\runauth{Author1 et al.}

\jid{nupha}

\jnltitlelogo{Nuclear Physics A}

\usepackage{graphicx}
\usepackage{amsmath,amssymb}
\usepackage{latexsym}
\usepackage{amsfonts}
\usepackage{color}
\usepackage{bbold}
\usepackage{bm}

\def\st{\begin{equation}}
\def\stp{\end{equation}}
\def\bg{\begin{eqnarray}}
\def\nd{\end{eqnarray}}

\def\x{{\bm x}}
\def\y{{\bm y}}

\def\mpt{\langle p_T \rangle}

\begin{document}

\begin{frontmatter}



\title{A scaling relation between proton-nucleus and nucleus-nucleus collisions}

\author{G\"ok\c ce Ba\c sar\footnote{Speaker}} 
\author{Derek Teaney}
\address{Department of Physics and Astronomy,  Stony Brook University, Stony Brook, NY 11794, USA}

\runauth{G\"ok\c ce Ba\c sar et al. }



\begin{abstract}
It is recently discovered that at high multiplicy, the proton-nucleus ($pA$) collisions give rise to two particle correlations that are strikingly similar to those of nucleus-nucleus ($AA$) collisions at the same multiplicity, although the system size is smaller in $pA$.  Using an independent cluster model and a simple conformal scaling argument, where the ratio of the mean free path to the system size stays constant at fixed multiplicity, we argue that flow in $pA$ emerges as a collective response to the fluctuations in the position of clusters, just like in $AA$ collisions. With several physically motivated and parameter free rescalings of the recent LHC data, we show that this simple model captures the essential physics of elliptic and triangular flow in $pA$ collisions. 
\end{abstract}

\end{frontmatter}


\section{Introduction}
As the recent measurements by the LHC \cite{CMS_flow,Abelev:2012ola,Aad:2012gla} and RHIC \cite{Adare:2013piz} collaborations have shown, the particle production in high multiplicity proton-nucleus ($pA$) collisions exhibits striking long range two particle correlations which are quantitatively similar to the corresponding correlator in nucleus-nucleus ($AA$) events at the same multiplicity. Some features of these correlations are reproduced by the Color Glass Condensate (CGC) without reference to the fluctuating geometry \cite{Dusling:2012cg,Dusling:2013oia,McLerran:2014uka,Gyulassy:2014lsa}. However, hydrodynamic simulations of $p+A$ events also qualitatively predicted the correlations observed in the data \cite{Bozek:2011if,Bozek:2012gr,Bzdak:2013zma}, suggesting that
the origin of the flow in $p+A$ is similar to $A+A$. The aim of this talk is to give a brief explanation for this similarity by arguing that both in high multiplicity $pA$ and in $AA$ events, such long range correlations emerge from a collective response to the underlying geometry. It turns out implementation of an independent cluster model, and a simple ``conformal scaling" framework \cite{Basar:2013hea} where the ratio of the mean free path to the system size is approximately the same for the $pA$ and $AA$ events are enough to capture the essential physics of such a collective response. According to this framework, at  a given multiplicity, the $pA$ event is smaller but hotter and denser, such that it develops a similar flow pattern as in $AA$. Below, these statements are made quantitative using both integrated and transverse momentum ($p_T$) dependent $v_2\{2\}$ and $v_3\{2\}$ measurements of LHC. It is also worth to mention that conformal scaling framework was applied to $p_T$ dependent $v_4\{2\}$ and $v_5\{2\}$ as well, giving excellent results\footnote{see talk by J. Jia at this conference}. All these findings provide a strong evidence for the existence of collective physics in $pA$ collisions. 
 
 \section{Independent cluster model and conformal dynamics}
 
We describe the initial state by $N_{clust}$ independently distributed clusters such that the multiplicity $N$ is proportional to $N_{clust}$. There is a single dimensionful parameter, say mean free path,  $l_{mfp}\propto T_i^{-1}$, in our model that controls the response dynamics. The conformal scaling is manifested by the assumption $l_{mfp}\,L^{-1}=f\left({dN\over dy}\right)$, where $L$ is the system size. For instance in a saturation inspired model the dimensionful parameter would be $Q_s$ where $N_{clust}=\pi Q_s^2L^2$ and $f=(dN/dy)^{-1/2}$. The distribution of clusters in the transverse plane, $n(\x)$, is random around their mean value, $\bar n(\x)$,
\begin{align}
n(\x)=\bar{n}(\x) + \delta n(\x)\quad,\quad\langle\delta n(\x)\delta n(\y)\rangle=\bar n(\x)\delta^{(2)}(\x-\y) 
\end{align}
 where $\delta n(\x)$ denotes the fluctuations around the average distribution \cite{Bhalerao:2006tp}. The flow emerges as a collective response to the geometry defined by the distribution of the clusters. We adopt linear response. The conformal scaling framework then dictates the response coefficients, $k_{2,3}, $ to depend only on $l_{mfp}/L$, hence the multiplicity, i.e.  $v_{2,3}=k_{2,3}(l_{mfp}/L)\epsilon_{2,3}$, where $\epsilon_2$ and $\epsilon_3$ are the eccentricity and triangularity respectively. 
 
It is important to understand what the sources of the eccentricity and triangularity are. The $AA$ events with multiplicity comparable to $pA$ events are peripheral and the eccentricity is sourced by \textit{both} the average cluster distribution and the fluctuations around it. On the other hand a high multiplicity $pA$ event is central and the eccentricity is sourced \textit{only} by the fluctuations.  The triangularity is sourced only by fluctuations both for $pA$ and $AA$.  A gaussian distribution for the clusters, which gives a very good approximation to more complicated Glauber models, leads to the following expressions for the mean $\epsilon_2^2$ and $\epsilon_3^2$
\begin{align}
\langle\epsilon_2^2\rangle_{AA}  =\epsilon_s^2  + \langle \delta \epsilon_2^2 \rangle,\quad \langle\epsilon_2^2\rangle_{pA}  =\langle \delta \epsilon_2^2 \rangle={\langle r^4\rangle\over N_{\rm clust}\langle r^2\rangle^2}\,\quad \langle\epsilon_3^2\rangle_{AA}  =\langle\epsilon_3^2\rangle_{pA}  =\langle \delta \epsilon_3^2 \rangle= {\langle r^6\rangle\over N_{\rm clust}\langle r^2\rangle^3}
 \end{align} 
where $\epsilon_s$ is the average eccentricity. Note that we have assumed the same transverse distribution for $pA$ and $AA$. In the first glance it seems like a dangerous assumption, but since what enter into the formulas above are double ratios, the sensitivity to different shapes is rather small (around $\sim 15\%$ at the most) and no fine tuning is needed.  

\section{Elliptic and triangular flow}

An immediate consequence of the discussion above is that at a given multiplicity, $v_3$ of $pA$ and $AA$ should be the same as they are both sourced by the fluctuations in the cluster distribution. As the LHC measurement shows, this is indeed true; $(v_3)_{pA}=(v_3)_{AA}$ up to a few percent \cite{CMS_flow}. In order to compare the $v_2$s justly, one needs to ``remove'' the effect of the average geometry and isolate the fluctuation driven part of $(v_2)_{AA}$, since in $pA$ fluctuations constitute the only source. This can be done by the following rescaling, where the scaling factor projects onto  $\sqrt{\langle\delta \epsilon_2^2\rangle}$:
\begin{align}
(v_2\{2\})_{\rm PbPb, rscl}\equiv\sqrt{1-{\epsilon_s^2\over\langle\epsilon_2^2\rangle_{PbPb}}} (v_2\{2\})_{\rm PbPb}.
\label{rescale}
\end{align}
\begin{figure}[h]
\center
\includegraphics[scale=0.45]{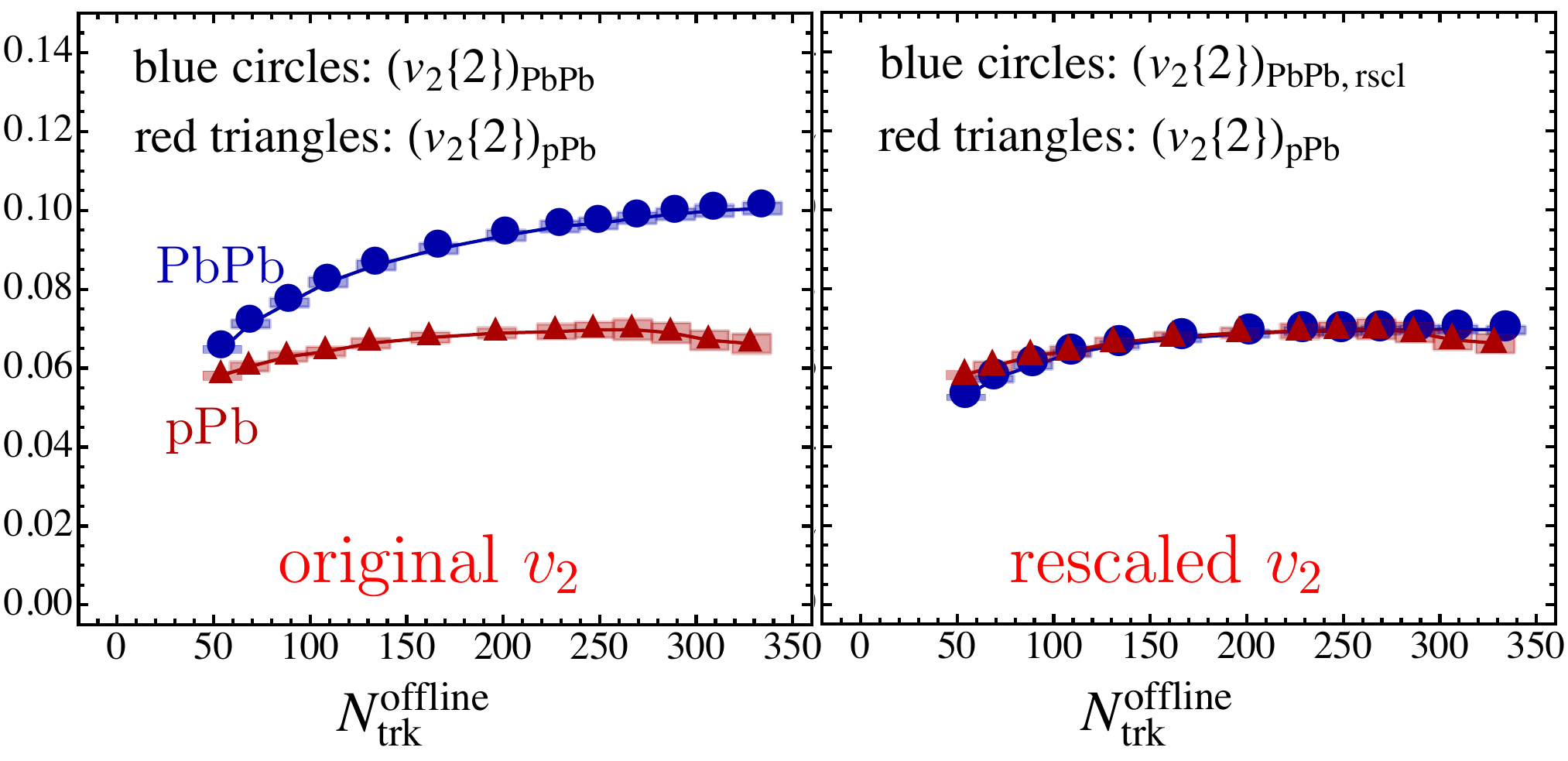}
\caption{The comparison between the fluctuation driven $v_2$ in $AA$ and $pA$. Left: The actual data. Right: The fluctuation driven part isolated in $AA$.}
\end{figure}

We have computed the rescaling factor in \eqref{rescale} via a Monte Carlo Glauber simulation. Once the average geometry is taken out, the conformal scaling predicts
\begin{align}
(v_2\{2\})_{\rm PbPb, rscl} \equiv k_2\sqrt{\langle\delta\epsilon_2^2\rangle_{PbPb}} \,\,\simeq\,\, k_2 \sqrt{\langle\delta.
\epsilon_2^2\rangle_{pPb}}\equiv(v_2\{2\})_{pPb}
\end{align}
The excellent agreement between the $v_2$s of $pA$ and $AA$ after the rescaling is shown in Figure 1. Note that there is no fitting parameter in the plot. Furthermore, the rescaling factor is a nontrivial function of centrality, hence multiplicity, and such a remarkable agreement is very nontrivial, indicating a strong evidence for a common origin for anisotropy in $pA$ and $AA$ which is a collective response to the geometry.    

Conformal dynamics also allows one to compare the transverse momentum, $p_T$, dependence of $v_2$ and $v_3$. Since $p_T$ is a dimensionful quantity, it should enter into the expression for $v_{2,3}(p_T)$ as
\begin{align}
v_{2} (p_T)= \epsilon_{2} \, f_{2}\left({p_T/ \mpt}\right) ,\quad v_{3} (p_T)= \epsilon_{3} \, f_{3}\left({p_T/ \mpt}\right) 
\end{align}
where the momentum dependent response coefficients $f_{2,3}$ are universal functions and the average transverse momentum $\mpt \sim l_{mfp}^{-1}\sim L^{-1}$ for fixed $dN/dy$. Note that such a relation is expected to hold for small $p_T\sim \mpt$, where we expect the collective behavior to dominate. As a consequence, in order to compare the momentum dependence of flow in $pA$ and $AA$, we should rescale the $p_T$ axis of the $AA$ to take into account the difference in $\mpt$ of $pA$ and $AA$. The prediction of the conformal scaling is then
\begin{align}
\left[v_2\{2\}(p_T)\right]_{pPb} =  \left[v_2\{2\}\left(p_T/\kappa\right)\right]_{PbPb,\rm rscl},\quad\left[v_3\{2\}(p_T)\right]_{pPb} =\left[v_3\{2\}\left(p_T/\kappa\right)\right]_{PbPb},\quad \kappa\equiv \mpt_{\rm pPb}/\mpt_{\rm PbPb}\simeq1.25\,.
\end{align}
The measurement of $\mpt$ is taken from \cite{ALICE_mean_pt}.  The original data for $v_2$ and $v_3$
 together with this complete (and parameter free) rescaling is shown in Figure 2.
 From the lower panels, we see that the agreement between the dimensionless slopes in the low $p_T$ region is remarkable, and seems to affirm the conformal rescaling. At higher $p_T$, the $v_2\{2\}$ starts to systematically differ. This difference seems to become larger for lower multiplicities where non-flow  could become significant.

 Another prediction of the conformal scaling is that, the ratio of the system sizes of $pA$ and $AA$ systems is roughly $L_{AA}/L_{pA}=\mpt_{pA}/\mpt_{AA}\approx1.25$. The recent Hanbury-Brown Twiss (HBT) measurement in LHC reveals that $R_{AA}/R_{pA}\approx1.4$ \cite{HBT}. Of course one expects some difference between the HBT radii and the system size as defined here, yet such an agreement is still remarkable. 

\begin{figure}[h]
\center
\includegraphics[scale=0.623]{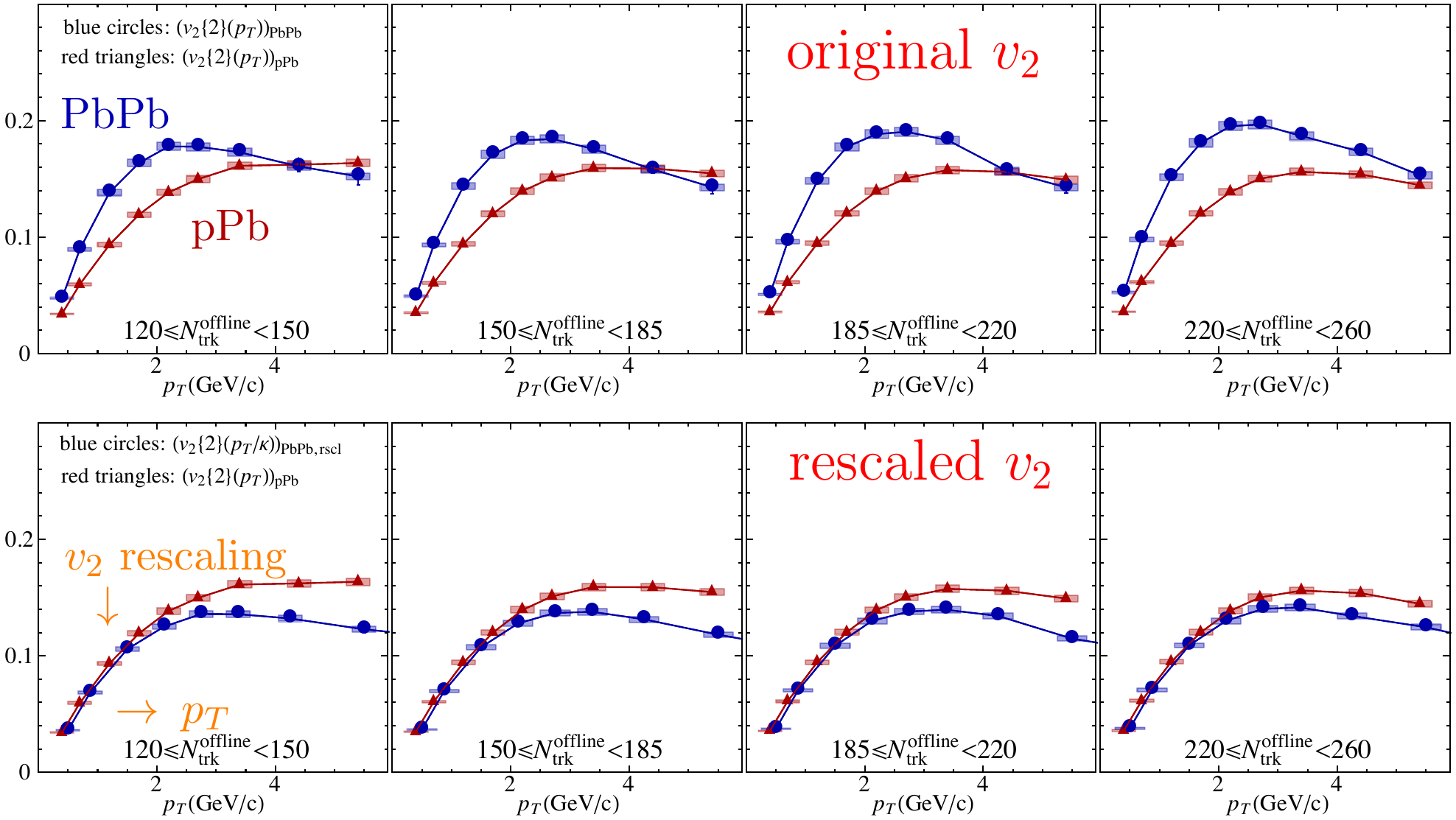}
\includegraphics[scale=0.655]{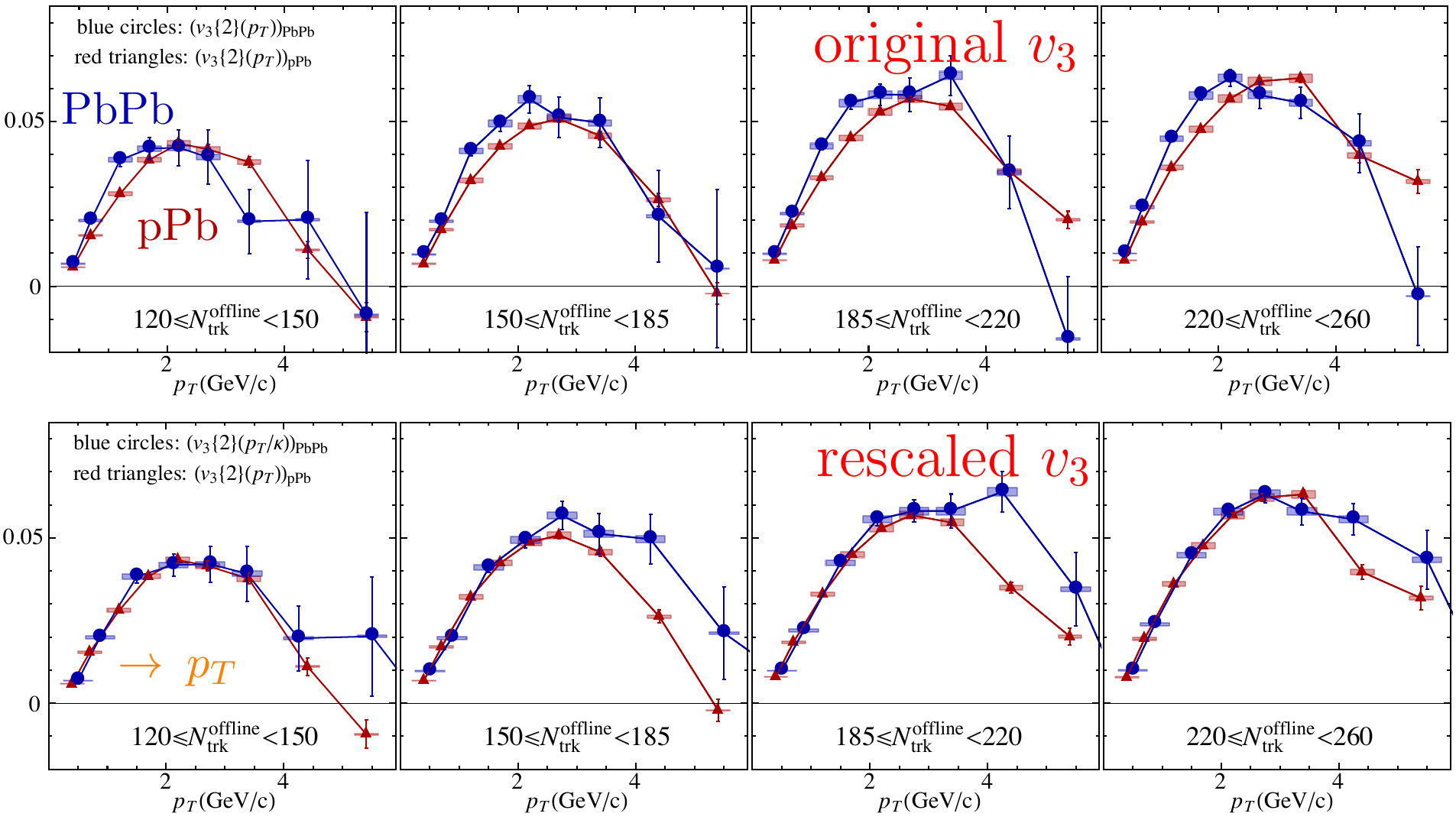}
\label{fig:v23pt}
\caption{Comparison of the momentum dependent $v_{2,3}$ for $pA$ and $AA$. The $p_T$ axes of $AA$  are rescaled by the ratio of $<p_T>$ of $pA$ to $AA$ as dictated by conformal dynamics. For $v_2$, the $y$-axis is also rescaled to isolate the fluctuation driven part.  The data is from \cite{CMS_flow}}
\end{figure}

\section{Conclusions}

The presented parameter free analysis of the two particle angular correlations in $pA$ and $AA$ collisions at the LHC, with several physically motivated rescalings based on a simple conformal scaling argument, provides an explanation for the similarity in two systems. First, once the effect of average geometry is taken from $AA$ measurement, the integrated $v_2\{2\}$ in $AA$ is the same as in $pA$ at fixed multiplicity.  The integrated $v_3\{2\}$ in these two colliding systems are already equal. Since the separation of $v_2$ into average and fluctuations in $AA$ was entirely motivated by linear response and geometry, it is reasonable to conclude that both the elliptic and triangular flow in $pA$ should also be understood as a linear response to initial geometric fluctuations.  Furthermore, the response coefficients of $pA$ and $AA$ are argued to be approximately equal based on the conformal scaling, which assumes a single dimensionful quantity controlling the response dynamics. The $p_T$ dependence of the $v_2$ and $v_3$ provide further support for such scaling under which both $v_2(p_T)$ and $v_3(p_T)$ curves for $AA$ collapse onto the ones for $pA$.  Consequently, phenomenologically, it seems highly unlikely that the angular correlations in $pA$ and $AA$ arise from different physics and likely that the underlying physics is collective response to fluctuation driven eccentricities.

This work was supported by the DOE under the grants DE-FG-88ER40388 (GB) and DE-FG-02-08ER4154 (DT).

\bibliographystyle{elsarticle-num}
\bibliography{pAbib}

\end{document}